# Cosmological gamma ray bursts and the highest energy cosmic rays

Eli Waxman

*Institute for Advanced Study, Princeton, NJ 08540; E-mail: waxman@guinness.ias.edu*

## Abstract

We discuss a scenario in which the highest energy cosmic rays (CR's) and cosmological $\gamma$-ray bursts (GRB's) have a common origin. This scenario is consistent with the observed CR flux above $10^{20}$eV, provided that each burst produces similar energies in $\gamma$-rays and in CR's above $10^{20}$eV. Protons may be accelerated by Fermi's mechanism to energies $\sim 10^{20}$eV in a dissipative, ultra-relativistic wind, with luminosity and Lorentz factor high enough to produce a GRB. For a homogeneous GRB distribution, this scenario predicts an isotropic, time-independent CR flux.

Three cosmic ray (CR) events with energies $> 10^{20}$eV have recently been observed by the Yakutsk [1], the Fly's Eye [2] and the Agasa [3] experiments. These events pose two challenges to current models for the production of CR's. First, the high energies rule out most of the acceleration mechanisms so far discussed [4]. Second, since the distance traveled by such particles must be smaller than 100Mpc [5], their arrival directions are inconsistent with the position of any astrophysical object that is likely to produce high energy particles [6].

The origin of gamma-ray bursts (GRB's) is yet unknown. However, recent observations give increasing evidence that GRB's originate from cosmological sources [7] (for a different point of view see, e.g., [8]), and a variety of plausible cosmological sources have been suggested [9]. Whatever the ultimate source is, observations strongly suggest the following similar scenario for the production of GRB's [10]. The rapid rise time, $\sim$ ms, observed in some bursts implies that the sources are compact, with a linear scale $r_0 \sim 10^7$cm. The high luminosity required for cosmological bursts, $\sim 10^{51}$ergs$^{-1}$, then results in an initially optically thick (to pair creation) plasma of photons, electrons and positrons, which expands and accelerates to relativistic velocities [11]. This is true whether the energy is released instantaneously, i.e. over a time scale $r_0/c$, as a "fireball", or as a wind over a duration comparable with the entire burst duration ($\sim$seconds). In fact, the hardness of the observed photon spectra, which extends to $\sim$ 100MeV, implies that the $\gamma$-ray emitting region must be moving relativistically, with a Lorentz factor of order a few hundreds [12].

If the observed radiation is due to photons escaping the fireball/wind as it becomes optically thin, two problems arise. First, the photon spectrum is quasi-thermal, in contrast with observations. Second, the plasma is expected to be "loaded" with baryons which may be injected with the radiation or present in the atmosphere surrounding the source. A small baryonic load, $\geq 10^{-8} M_{\odot}$, increases the optical depth (due to Thomson scattering) so that most of the radiation energy is converted to kinetic energy of the relativistically expanding baryons before the plasma becomes optically thin [13]. To overcome both problems Rees & Mészáros [14] proposed that the observed burst is produced once the kinetic energy of the ultra-relativistic ejecta is re-randomized by some dissipation process at large radius, beyond the Thomson photosphere, and then radiated as $\gamma$-rays. Collision of the relativistic baryons with the inter-stellar medium [14], and internal collisions within the ejecta itself [15,16], were proposed as possible dissipation processes.

In this *letter* we consider the possibility that high energy CR's and cosmological GRB's are related phenomena. We find that the observed flux of CR's beyond $10^{20}$eV is consistent with a scenario in which these particles are produced in GRB's provided that each burst produces similar energies in $\gamma$-rays and highest energy CR's. Adopting a dissipative ultra-relativistic wind model for the production of GRB's, we then demonstrate that protons may be accelerated in such a wind to energies of order $10^{20}$eV by Fermi's mechanism [17]. These two remarkable coincidences possibly suggest that GRB's and highest energy CR's have a common origin.

*GRB rate and CR flux.* The high energy CR experiments observe a cone around the zenith of opening angle $\sim 45^o$ over a period of $\sim$ 10yr. The rate of cosmological GRB events is $\nu_\gamma \sim 3 \times 10^{-8}\text{Mpc}^{-3}\text{yr}^{-1}$ [18]. The rate of events in a cone observed by CR experiments extending to a distance of $\sim$ 100Mpc, which is the maximal distance likely to be traveled by a CR of $10^{20}$eV, is $\sim$ 1 per 50yr. The rate in a cone extending to 50Mpc, the maximal distance traveled by the $3 \times 10^{20}$eV CR observed by the Fly's Eye, is 8 times smaller. Thus, the probability of the experiments to observe a CR pulse from a GRB event over $\sim$ 10yr period is small, unless the CR pulse is broadened in time, due to propagation through the inter-galactic medium (IGM), over a time scale $\geq$ 100yr.

Time broadening is likely to occur due to the combined effects of deflection by random magnetic fields and energy dispersion of the particles. Consider a proton of energy $E$ propagating through a magnetic field of strength $B$ and correlation length $\lambda$. As it travels a distance $\lambda$, the proton is typically deflected by an angle $\alpha \sim \lambda/R_L$, where $R_L = E/eB$ is the Larmor radius. The typical deflection angle for propagation over a distance $d$ is $\theta_s \sim (d/\lambda)^{1/2}\lambda/R_L$. This deflection results in a time delay, compared to propagation along a straight line, of order

$$\tau(E) \approx \theta_s^2 d/c \approx \left(\frac{eB}{E}d\right)^2 \frac{\lambda}{c}. \tag{1}$$

((1) is valid for small $\theta_s$). The time delay is sensitive to the particle energy. Thus, a spread in particle energies $\Delta E \sim E$ results in a time broadening of the pulse over a time $\sim \tau(E)$. Since the proton energy loss is a random process (collisions with the microwave background), there is a large dispersion in particle energies for a propagation distance over which significant energy loss occurs. For protons with energies $> 10^{20}$eV the RMS energy spread is comparable to the average energy over propagation distances in the range $10 - 100$Mpc [5]. Thus, the CR pulse would be spread over a time $\tau(E)$. The field required to produce $\tau > 100$yr is

$$B > 2 \times 10^{-12} E_{20} d_{100}^{-1} \lambda_{10}^{-1/2}\,\mathrm{G}, \tag{2}$$

where $E = 10^{20} E_{20}$eV, $d = 100 d_{100}$Mpc and $\lambda = 10\lambda_{10}$Mpc. The required field is not inconsistent with our knowledge of IGM properties. Estimates of the average (over cosmological scales) IGM field are of order $3 \times 10^{-11}$ [19], while disordered fields may be larger. A time broadening over $\tau > 100$yr is therefore reasonable.

The time variability and directional properties of CR flux produced by GRB events would depend on the distribution of GRB's in our vicinity and on the structure of the IGM magnetic fields. A detailed analysis of the flux properties are therefore beyond the scope of this *letter*. However, an order of magnitude estimate of the required CR production by a GRB may be obtained as follows. If GRB events are homogeneously distributed over scales of order $\sim 100$Mpc, and if $\tau \gg 100$yr, then the CR flux due to GRB's would be approximately isotropic and time independent. The observed CR flux at energies beyond $10^{20}$eV is $\sim 3 \times 10^{-17}\,\mathrm{m}^{-2}\mathrm{s}^{-1}\mathrm{sr}^{-1}$ [20], corresponding (for isotropic time independent flux) to a number density $n_{CR} \sim 10^{-30}\mathrm{cm}^{-3}$ of these particles. If this density is produced by GRB's then each GRB event should produce $N = n_{CR}/\nu_\gamma \tau_{CR}$ protons beyond $10^{20}$eV, where $\tau_{CR} \approx 3 \times 10^8$yr is the life time of protons with energies $> 10^{20}$eV [5]. The energy in protons beyond $10^{20}$eV, that should be released in each GRB in order to be consistent with the observed CR flux is therefore

$$E_{CR} \approx 5 \times 10^{50} \bar{E}_{20} \mathrm{erg}. \tag{3}$$

Here, $\bar{E} = 10^{20} \bar{E}_{20}$eV is the average energy of protons beyond $10^{20}$eV. Thus, $E_{CR}$ is required to be similar to the energy emitted as $\gamma$-rays by a cosmological GRB, $E_\gamma \approx 2 \times 10^{51}$erg.

We have so far assumed that the photon emission from GRB's is spherically symmetric. If the emission is concentrated into a solid angle $\delta\Omega$, then the rate of GRB events is higher by a factor $4\pi/\delta\Omega$ and the $\gamma$-ray energy produced by each event is smaller by the same factor. The above conclusions remain unchanged, except that $E_{CR}$ in (3) is also reduced by a factor $4\pi/\delta\Omega$.

*Fermi acceleration in dissipative wind models of GRB's.* We consider a compact source producing a wind, characterized by an average luminosity and mass loss rate, denoted $L$ and $\dot{M} = L/\eta c^2$ respectively. At small radius, the wind bulk Lorentz factor, $\gamma$, grows linearly with radius, until most of the wind energy is converted to kinetic energy and $\gamma$ saturates at $\gamma \sim \eta$. We assume that the wind is ultra-relativistic, $\gamma \gg 1$, and that internal shocks in the ejecta reconvert a substantial part of the kinetic energy to internal energy at $r \sim r_d$. Since $\gamma \gg 1$, substantial dissipation of kinetic energy implies that the random motions in the wind rest frame are relativistic. The relativistic random motions are likely to give rise to a turbulent build up of magnetic fields, and therefore to second order Fermi acceleration of charged particles. We derive below the constraints that should be satisfied by the wind parameters in order to allow acceleration of protons to $\sim 10^{20}$eV.

The most restrictive requirement, which rules out the possibility of accelerating particles to energies $\sim 10^{20}$eV in most astrophysical objects, is that the particle Larmor radius $R_L$ should be smaller than the system size [4]. In our scenario we must apply a more stringent requirement, namely that $R_L$ should be smaller than the largest scale $R$ over which the magnetic field fluctuates, since otherwise the second order Fermi acceleration may not be efficient. We may estimate $R$ as follows. The comoving time, i.e. the time measured in the wind rest frame, is $t = r/\gamma c$. Thus, at $r = r_d$ regions separated by a comoving distance larger than $r_d/\gamma$ are causally disconnected, and the wind properties fluctuate over comoving length scales up to $R \sim r_d/\gamma$. We must therefore require $R_L < r_d/\gamma$. A somewhat more stringent requirement is related to the wind expansion. Due to expansion the internal energy is decreasing and therefore available for CR acceleration (as well as for $\gamma$-ray production) only over a comoving time $t_d \sim r_d/\gamma c$. The typical Fermi acceleration time is $t_a = fR_L/c\beta^2$, where $\beta c$ is the Alfvén velocity and $f \sim$ few [21,4]. In our scenario $\beta \sim 1$ leading to the requirement $fR_L < r_d/\gamma$. This condition sets a lower limit for the required comoving magnetic field strength. In obtaining this limit one should note, that due to the relativistic motion the energy of the accelerated particle is $\sim \gamma$ times smaller in the wind rest frame than in the observer frame. Thus, the Larmor radius is given by $R_L = E/\gamma eB$, where $E$ is the observed particle energy. Using this expression for $R_L$ we obtain

$$B > f\frac{E}{er_d} = 3 \times 10^4 f E_{20} r_{d,13}^{-1}\mathrm{G}, \tag{4a}$$

where $r_d = 10^{13} r_{d,13}\mathrm{cm}$. 4a can be stated in a different form by comparing the required field to the equipartition field, i.e. to a field with comoving energy density similar to that associated with the random energy of the baryons. Substantial dissipation of kinetic energy implies that the contribution of the random motion energy density to the luminosity is large. In this case (4a) may be put in the form

$$\left(\frac{B}{B_{e.p.}}\right)^2 > 0.15 f^2 \gamma_{300}^2 E_{20}^2 L_{51}^{-1}, \tag{4b}$$

where $B_{e.p.}$ is the equipartition field, $\gamma = 300\gamma_{300}$, and $L = 10^{51} L_{51}\mathrm{ergs}^{-1}$.

The accelerated proton energy is also limited by energy loss due to synchrotron radiation and interaction with the wind photons. The condition that the synchrotron loss time, $t_{sy} = (6\pi m_p^4 c^3/\sigma_T m_e^2)E^{-1}B^{-2}$, should be smaller than the acceleration time $t_a$ is

$$B < 3 \times 10^5 f^{-1} \gamma_{300}^2 E_{20}^{-2}\mathrm{G}. \tag{5}$$

This condition may be satisfied simultaneously with (4a) provided that the dissipation radius is large enough, i.e.

$$r_d > 10^{12} f^2 \gamma_{300}^{-2} E_{20}^3 \mathrm{cm}. \tag{6}$$

The high energy protons we are interested in lose energy in interaction with the wind photons mainly through collisions producing pions, for which $\sim 10\%$ of the energy is lost per collision [5]. The typical time for energy loss in this process is $t_\gamma \sim 10/n_\gamma \sigma c$, where $\sigma \approx 10^{-29}\text{cm}^2$ is the collision cross section for high energy protons and $n_\gamma$ is the photon number density. $n_\gamma$ is approximately given by $L_\gamma = 4\pi r_c^2 c\gamma n_\gamma \varepsilon_\gamma$, where $\varepsilon_\gamma$ is the observed photon energy, $\varepsilon_\gamma \sim$ MeV, and $L_\gamma$ is the gamma-ray luminosity. The condition $t_\gamma > t_a$ gives

$$B > 20 f L_{\gamma,51} E_{20} r_{d,13}^{-2} \gamma_{300}^{-2} \text{G}. \tag{7}$$

Let us discuss the implications of (4-7) for a wind which should produce a GRB. The bulk Lorentz factor of the wind at $r_d$ must satisfy $\gamma_{300} \geq 1$. If the process converting the random energy to gamma-rays is not very inefficient, then $L$ is a few times higher than the observed gamma-ray luminosity, $L_\gamma \approx 10^{51}\text{ergs}^{-1}$. (4b) then implies that the magnetic field energy should be close to equipartition. This requirement is independent of the value of $r_d$ and determined only by the wind luminosity and Lorentz factor. The condition (5) is satisfied in this case provided that dissipation occurs at large enough radius, as required by (6) (this also ensures that (7) is satisfied). In the dissipative wind models for cosmological GRB's, proposed by Paczyński & Xu [15] and by Rees & Mészáros [16], internal shocks occur due to time variability of the source. Variability that leads to fluctuations in the bulk Lorentz factor $\Delta\gamma \sim \gamma$ over a time scale $\Delta t$, produces internal shocks which dissipate a substantial fraction of the wind kinetic energy at $r_d \approx \gamma^2 c\Delta t$. The flux emitted by the source may fluctuate on time scales ranging from the source dynamical time, $t_{dyn} \sim r_0/c \sim 10^{-3}$s, to the wind duration $T \sim$ s, leading to dissipation radii $r_d$ in the range $3\times 10^{12}\gamma_{300}^2$cm to $3\times 10^{15}\gamma_{300}^2$cm. The dissipation radii in these models may therefore be large enough to satisfy (6) and allow the production of high energy protons.

The dissipation radius of a GRB producing wind is limited by the observed duration, $T \sim$ s, of the $\gamma$-ray pulse, $r_d < \gamma^2 cT$. This condition may be satisfied simultaneously with (6) provided that the Lorentz factor satisfies

$$\gamma > 40 f^{1/2} E_{20}^{3/4} T_1^{-1/4}, \tag{8}$$

where $T = T_1$s.

The estimates given above were obtained assuming that the wind is spherically symmetric. However, since the wind is ultra-relativistic, the behavior of a jet-like wind is similar to that of a spherical one, provided that the jet opening angle is larger than $1/\gamma$. In this case, $L$ should be understood as the luminosity that would have been produced by the wind if it were spherically symmetric. A possible complication that may arise for jet-like wind is that particles may escape the acceleration region transversally on a time scale shorter than $t_a$ (the condition $R_L < r_d/\gamma$ ensures that the radial escape time is larger than $t_a$). However, for jets with opening angles $> 1/\gamma$ the transversal size of the jet is larger than $r_d/\gamma$, so that the transversal escape time is also larger than $t_a$.

In conclusion, we have found that the observed flux of CR's with energies $> 10^{20}$eV is consistent with a scenario in which these particles are produced in cosmological GRB's provided that each burst produces similar energies in gamma-rays and CR's beyond $10^{20}$eV. We have also demonstrated that a dissipative ultra-relativistic wind, with luminosity and bulk Lorentz factor implied by GRB observations, satisfies the constraints necessary to allow the acceleration of protons to energies $\sim 10^{20}$eV by second order Fermi acceleration. We emphasize that this energy estimate is independent of the wind model details, and depends only on the wind luminosity and average Lorentz factor, which are directly estimated from observations (assuming a cosmological origin for GRB's).

Acceleration to $\sim 10^{20}$eV is possible provided that the dissipation of kinetic energy occurs at radii $r_d \geq 10^{13}$cm [(6)], and that turbulent build up of magnetic fields during dissipation gives rise to fields which are close to equipartition [(4b)]. Although the details of the mechanism of turbulent build up of an equipartition field are not fully understood, it seems to operate in a variety of astrophysical systems. We note that a magnetic field of this strength may exist in the plasma prior to the onset of internal collisions if a substantial part of the wind luminosity is initially, i.e. at $r \sim r_0$, provided by magnetic field energy density, and if the field becomes predominantly transverse. The pre-existing field may suppress small scale turbulent motions [16]. In this case shocks coherent over a scale $R \sim r_d/\gamma$ would exist and protons would be accelerated by first rather than second order Fermi mechanism. The constraints (4-7) are valid in this case too, therefore leaving the above conclusions unchanged.

Future investigations of a scenario, in which the highest energy CR's are produced by cosmological GRB's, should aim at obtaining characteristic signatures of such a scenario, which may be compared with observations. A homogeneous distribution of GRB events over scales $\sim$ 100Mpc would naturally lead to isotropic time independent CR flux. This prediction may be compared to observations once larger high energy CR detectors are operative. The CR energy spectrum, that should be produced by GRB's, is already constrained by observations. The life time of protons with energies $> 10^{19}$eV is larger than the Hubble time, $\sim 10^{10}$yr, which is $\sim 30$ times the life time of protons above $10^{20}$eV ($\sim 3 \times 10^{8}$yr). Thus, if the number of protons produced above $10^{19}$eV in a GRB event is similar to the number of protons produced above $10^{20}$eV, then the ratio of the flux above $10^{19}$eV to the flux above $10^{20}$eV would be $\sim 30$. The observed ratio is $\sim 300$. Thus, if the number of CR's produced per energy interval is a power law $dN \propto E^{-n}dE$, as typically expected from Fermi acceleration [4], we must have $n \leq 2$. A more detailed analysis of the acceleration process, resulting in a prediction for the produced energy spectrum, may therefore be directly compared with observational constraints. Note, that if the power law spectrum extends from $10^{21}$eV down to $10^{12}$eV (the lower bound is due to the fact that the source bulk Lorentz factor is $\sim 10^{3}$), then $n \leq 2$ implies that the total energy of relativistic protons is $\leq 5 \times 10^{51}$erg [cf. (3)]. Since the observed gamma-ray energy is $\approx 2 \times 10^{51}$erg, we do not expect the energy associated with CR's to change significantly the constraint on the total GRB source energy.

I thank J. W. Cronin for introducing me to the subject of highest energy cosmic rays, J. N. Bahcall and R. M. Kulsrud for invaluable comments, and F. Rasio and J. Goodman for helpful discussions. This research was partially supported by a W. M. Keck Foundation grant and NSF grant PHY 92-45317.